\documentclass[%
reprint,
 unsortedaddress,
 amsmath,amssymb,
 aps,prl
]{revtex4-1}

\usepackage{amsmath,amssymb}
\usepackage{euscript} 
\usepackage{mathptmx} 
\usepackage{ulem} 
\usepackage{graphicx}
\usepackage{dcolumn}
\usepackage{braket}
\usepackage[usenames, dvipsnames]{color}
\usepackage{float}
\usepackage{hyperref}
\usepackage[mathlines]{lineno}

\begin{document}

\preprint{APS/123-QED}

\title{Non-conventional topological band properties and gapless helical edge states in \\ elastic phononic waveguides with Kekul\'e distortion}

\author{Ting-Wei Liu}
\author{Fabio Semperlotti}
 \email{fsemperl@purdue.edu}
\affiliation{Ray W. Herrick Laboratories, School of Mechanical Engineering, Purdue University, West Lafayette, Indiana 47907, USA.}

\date{\today}

\begin{abstract}
This study investigates the topological behavior of a continuous elastic phononic structure characterized by a 3-term Kekul\'e distortion. The elastic waveguide consists of a hexagonal unit cell whose geometric dimensions are intentionally perturbed according to a generalized Kekul\'e scheme. The resulting structure exhibits an effective Hamiltonian that resembles a quantum spin Hall system, hence suggesting that the waveguide can support helical topological edge states.
Using first-principles calculations, we show the existence of nondegenerate pseudospin states and a very peculiar 6-lobe pseudospin texture and Berry curvature pattern.
Important insights are also provided concerning the topological states in the Kekul\'e lattice, so far considered indistinguishable, and their critical role in enabling unique gapless edge states, typically not achievable in phononic systems.
\end{abstract}

\maketitle
In the wake of the discovery of topological states of matter in quantum mechanical systems, researchers have explored the possibility to create analogue effect in classical waveguide systems including electromagnetic, acoustic, and elastic systems. In all these systems, the common interest lied in the possibility to develop scattering-free waveguides achieving, ideally, maximum transmission properties.

In acoustic and elastic systems, most of the work conducted in recent years has concentrated on creating mechanisms analogue to the Hall effect in its different forms. Systems preserving time-reversal (\textit{T}-) symmetry have been of particular interest owing to their manufacturing and operating simplicity. Examples include the acoustic analogue of the quantum valley Hall effect (QVHE) \cite{ValleySonicBulk,ValleySonicEdge,pal2017edge,vila2017observation,myqv,QVEXP,diatom,yan2018chip,snowflake,ma2019valley}
and of the quantum spin Hall effect (QSHE) \cite{MechanicalTI,AcousticTIPlate,AcousticTIAir,fold2017,
acousticTI2017,zonefold2,kekuleprl2017,kekulenjp2018,zonefold2018,
xia2018programmable,miniaci2018experimental,zhang2017topological,
yves2017topological,deng2017observation}. The later cases, also referred to as the acoustic topological insulator (TI), is achieved by leveraging pseudospins and degenerate double Dirac cones in mechanical lattices that allow mapping the effective Hamiltonian of the system to that of QSHE.
A comparison between systems utilizing the two \textit{T}-symmetry preserving mechanisms is investigated by Deng, \textit{et al.} \cite{vs}.
Moreover, Paulose \textit{et al.} \cite{dislocation} showed that topological zero-frequency soft modes can exist near the local dislocated deformation of a Kagome or square lattice. We note that this study did not report any occurrence of topological edge states along the edge dislocation.

Recently Liu \textit{et al.} \cite{kekuleprl2017} and Zhou \textit{et al.} \cite{kekulenjp2018} proposed discrete spring-mass models of phononic lattices in which the stiffness of the springs forming the unit cell was varied according to the Kekul\'e distortion pattern, as a way to devise QSHE analogue. In those studies, a perturbation approach was used to predict the formation of globally degenerate pseudospin modes. Note that these models can only capture the effective behavior in the vicinity of the degeneracy (i.e. the Dirac point) and are valid only for small values of the Kekul\'e's distortion strength.

In this paper, we propose a continuous elastic phononic waveguide based on a 3-term Kekul\'e texture which introduces a component breaking space-inversion (or parity \textit{P}-) symmetry.
We first derive the effective Hamiltonian near the Dirac point by using low-order perturbation theory. The analysis of the Hamiltonian will show that such waveguide exhibits a dynamic behavior that is indeed the analogue of QSHE.
Further, by using first-principle numerical calculations,
we reveal \textit{finer structures} of the phononic bands, namely: 1) the spurious degeneracy (predicted by low-order perturbation model) of the pseudospin states is lifted, and 2) the existence of an unusual six-lobe pseudospin texture and Berry curvature distributions around the $\Gamma$ point.
These new fundamental findings could not be observed based on the conventional 4-level Hamiltonian model and provide a critical foundation for the analysis of the pseudospin-resolved Berry phase. In addition, some topological ambiguities, including the gauge dependence in mapping the system dynamics to the QSHE, are also revealed and extensively discussed. Then, we
perform a thorough analysis of possible edge states supported by various domain wall configurations. These numerical results serve as a basis to clarify the nature of topological  states in Kekul\'e lattices so far considered indistinguishable. This important insight allows elucidating the mechanism leading to the existence of gapless edge states. 
In fact, in contrast with existing acoustic topological materials based on zone-folding strategy \cite{MechanicalTI, AcousticTIAir, fold2017, acousticTI2017, zonefold2, kekuleprl2017, kekulenjp2018, zonefold2018, xia2018programmable, zhang2017topological, yves2017topological, deng2017observation}, the generalized Kekul\'e distortion approach provides an additional degree of freedom that opens the way to realizing strictly gapless edge states (hence, further enhancing the robustness against back-scattering).
Another very interesting and unique consequence of the above mentioned gauge dependence is the existence of edge states on edge dislocations of the Kekul\'e lattice.

\begin{figure}[h]
\includegraphics[scale=0.41]{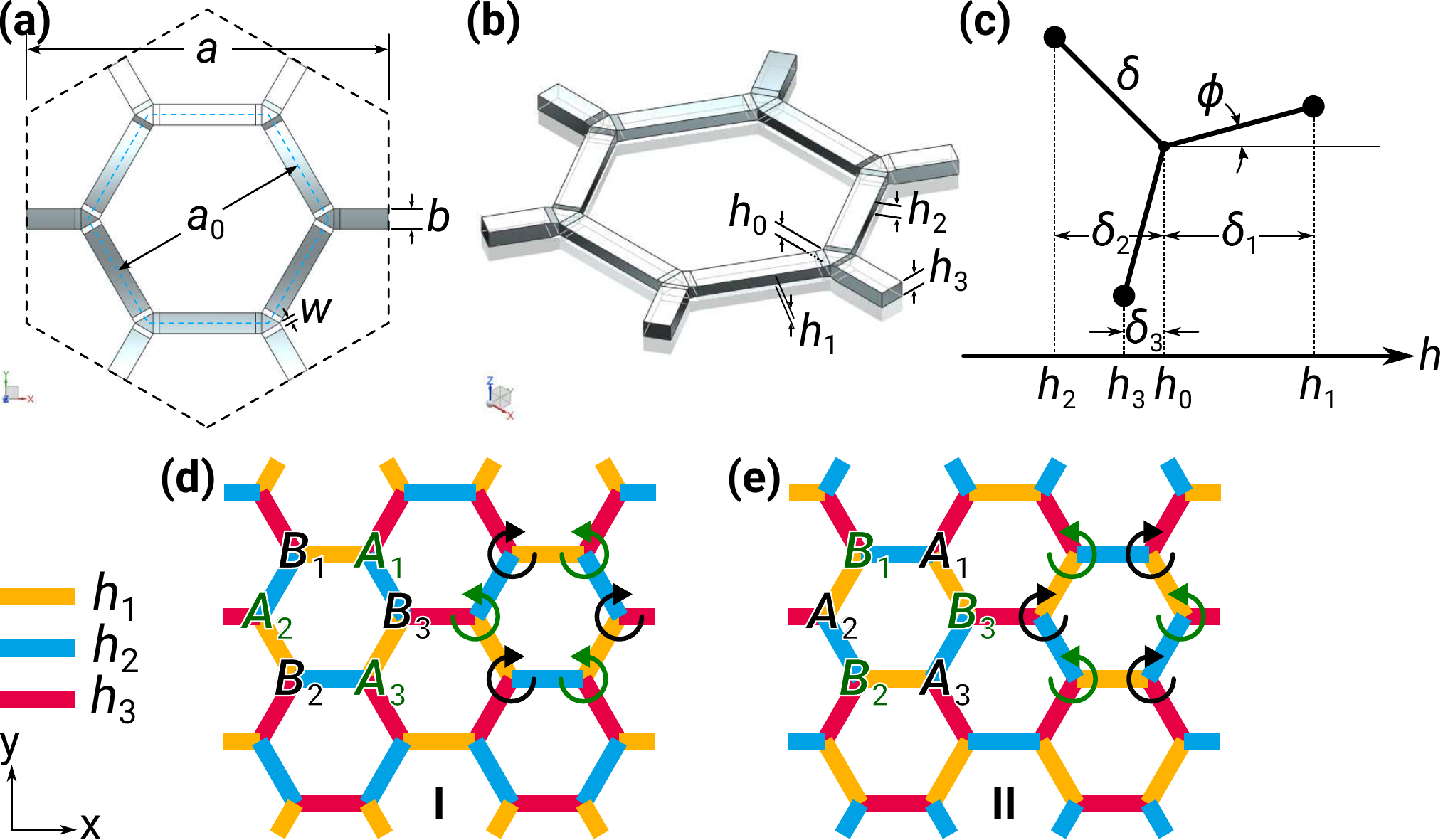}
\caption{\label{fig:lattice} (Color online) Schematic illustrations of the lattice geometry. (a) Top view of a primitive unit cell. (b) Isometric view of the unit cell indicating the thickness of the Kekul\'e pattern. (c) A graphical illustration showing the relations between the geometric parameters $\phi$, $\delta_{1,2,3}$, and $h_{1,2,3}$. (d) Illustration of type-I lattice, in which the $A$ ($B$) prisms are connected to beams with height $h_{1,2,3}$ in a counter-clockwise (clockwise) permutation. (e) Illustration of type-II lattice, where the permutations are inverted with respect to type-I.
}
\end{figure}

The material system consists in a phononic waveguide made of a reticular plate having a hexagonal lattice structure (Fig. \ref{fig:lattice}(a)) with constant $a$. Each slender beam in the reticular structure has a rectangular cross-section with width $b$ and varying height.
The Kekul\'e distortion pattern is imprinted on the phononic lattice by controlling the height ($h_{i}$) of each connecting beam. Beams with three different heights are connected to a triangular prism having average height $h_0$ (Fig.~\ref{fig:lattice}(b)). Their height varies linearly in a small section $w$ connected to the prism (Fig.~\ref{fig:lattice}(a)) and their distribution is symmetric with respect to the mid-plane of the lattice.

The classic Kekul\'e pattern has two different bonds linking the atoms \cite{kekule1865}. Here a generalized Kekul\'e pattern is considered, where all three beams connected to the prism have different height expressed by
\begin{equation} \label{eq:dist}
\begin{aligned} 
h_i &= h_0 + \delta_i,\\
\delta_i &= \delta \cos\Big(\phi + (i-1)\frac{2\pi}{3}\Big), \quad i = 1,2,3
\end{aligned}
\end{equation}
The perturbation typical of the Kekul\'e pattern can be graphically interpreted as in Fig.~\ref{fig:lattice}(c) whereas the abscissa indicates the height $h$. The three parameters (average height $h_0$, distortion amplitude $\delta$, and initial phase $\phi$) can construct any set of three real numbers $h_{1,2,3}$. For the numerical simulations reported in this paper, we set $h_0=a_0/20$, $\delta=h_0/4$, $b=a_0/10$, $w=a/60$, and chose aluminum as base material. With these assumptions, the only free parameter is $\phi$. Given an input value of $\phi$, two types of lattices can be created based on different cyclic permutation orders of $h_{1,2,3}$ around a specific prism. The lattices are labeled as type-I and type-II, as illustrated in Fig. \ref{fig:lattice} (d,e). In type-I lattice, $h_{1,2,3}$ follows a counter-clockwise permutation around the $A$ prisms and clockwise around the $B$ prisms, and vice versa in type-II lattice.

Note that a substitution $\phi \rightarrow \phi+2\pi/3$ results in $h_{1,2,3} \rightarrow h_{2,3,1}$ which does not change the permutation order. This transformation produces a rigid translation $A_3 \rightarrow A_1$ (or $\Vec{T}=(0,a_0)$) of the entire lattice therefore having no effect on the bulk  properties. However, we anticipate that when finite boundaries or interfaces (such as domain walls) are considered, this translation can produce non negligible effects. More details will be provided below.

\begin{figure}[ht]
\includegraphics[scale=.48]{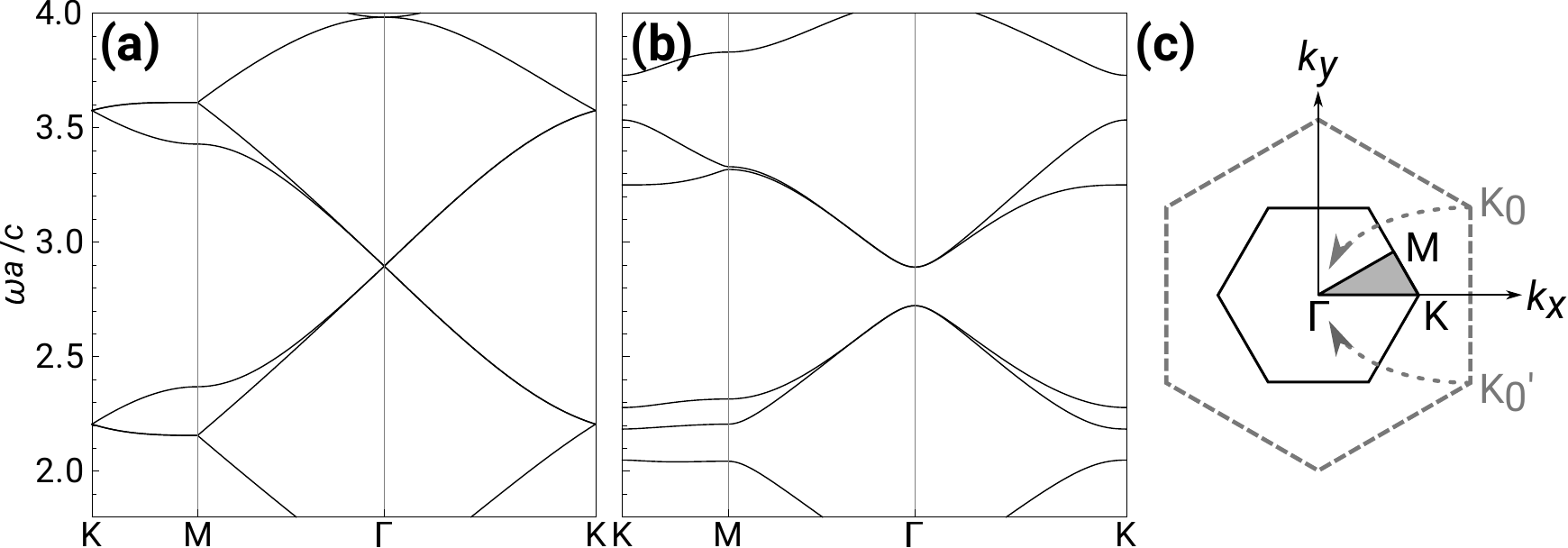}
\caption{\label{fig:bs} Phononic band structures of (a) the undistorted lattice and (b) the lattice with Kekul\'e distortion. (c) The zone folding illustration: the valleys of the original lattice overlap at $\Gamma$.}
\end{figure}

When $\delta=0$, hence in the absence of Kekul\'e distortion, the lattice has $C_{6v}$ symmetry and the unit cell is a $\sqrt{3}$-sized superlattice of its primitive cell with a corresponding folding of the original Brillouin zone. Following this folding, the Dirac cones at the original valleys $K_0$ and $K'_0$ overlap at the $\Gamma$ point (see Fig.~\ref{fig:bs} (c)), forming a double cone having 4-fold degeneracy (Fig.~\ref{fig:bs} (a)).
When the distortion is introduced, it breaks the original lattice periodicity (having lattice constant $a_0$ shown in Fig. \ref{fig:lattice} (a)) and lowers the symmetry to $C_{3v}$. Note that, differently from all previously considered acoustic zone-folding QSHE implementations \cite{acousticTI2017,zonefold2,zonefold2018,AcousticTIAir}, the Kekul\'e pattern breaks \textit{P}-symmetry. Once the Kekul\'e distortion is applied, the original 4-fold degeneracy at $\Gamma$ is lifted (due to mixing of the two valley modes) and a bandgap opens between two 2-fold degenerate bands, as shown in Fig.~\ref{fig:bs} (b). This specific example considers $\phi=30^\circ$ which yields $h_1 > h_3 > h_2$.

In order to understand the general characteristics that could be expected in such bulk medium, we derive the effective Hamiltonian which describes the eigenstates near the $\Gamma$ point of the distorted lattice. We assume that these states can be expanded using the 4-fold degenerate modes of the undistorted lattice as basis. Since the states of the distorted and undistorted lattices are defined in different spaces (different lattice geometry), a homeomorphic transformation is required to map the eigenstates of the distorted lattice back to the undistorted one. Once the eigenstates are expressed under the same frame, their inner product can be readily performed and used to the determine the expansion coefficients. The 4-fold degenerate basis contains two $g$- and two $h-$ orbit-like (with eight and ten angular nodal planes) states, respectively. A pseudospin basis can be constructed by generating linear combinations of symmetry adapted states as follows
\begin{equation}\label{eq:pseudospin}
\begin{aligned}
\ket{g\uparrow}&= \frac{1}{\sqrt{2}}(\ket{g_1}+i\ket{g_2}),~
\ket{g\downarrow}=\frac{1}{\sqrt{2}}(\ket{g_1}-i\ket{g_2}),\\
\ket{h\uparrow}&=\frac{1}{\sqrt{2}}(\ket{h_1}+i\ket{h_2}),~
\ket{h\downarrow}=\frac{1}{\sqrt{2}}(\ket{h_1}-i\ket{h_2}).
\end{aligned}
\end{equation}
(See \cite{Supp} for a video of the eigenstates). With the basis in the order $(\ket{g\uparrow},\ket{h\uparrow},\ket{g\downarrow},\ket{h\downarrow})$, the $4\times 4$ effective Hamiltonian (complete derivation in \cite{Supp}) assume a form similar to the Bernevig-Hughes-Zhang model \cite{bhz} 
\begin{equation}\label{bhz}
   H =
   \begin{pmatrix}
   h(\mathbf{k}) & 0 \\
   0 & h^*(\mathbf{-k})
   \end{pmatrix},
\end{equation}
in which $ h(\mathbf{k}) = (C-Dk^2)\sigma_0 + \EuScript{A}(k_x\sigma_x-k_y\sigma_y) + (M+Bk^2)\sigma_z$, where $\sigma_{x,y,z}$ are Pauli matrices, and $\sigma_0$ is the identity matrix.
$M$ indicates the coupling of $\ket{g}$ and $\ket{h}$ states yielding the bandgap. Also, recalling that the spin Chern number $C_S=\pm (sgn(M)+sgn(B))/2$,  the same sign of $M$ and $B$ will yield $C_S = \pm 1$ hence suggesting the existence of edge states at the interface of topologically distinct materials.

\begin{figure}[ht]
\includegraphics[scale=0.76]{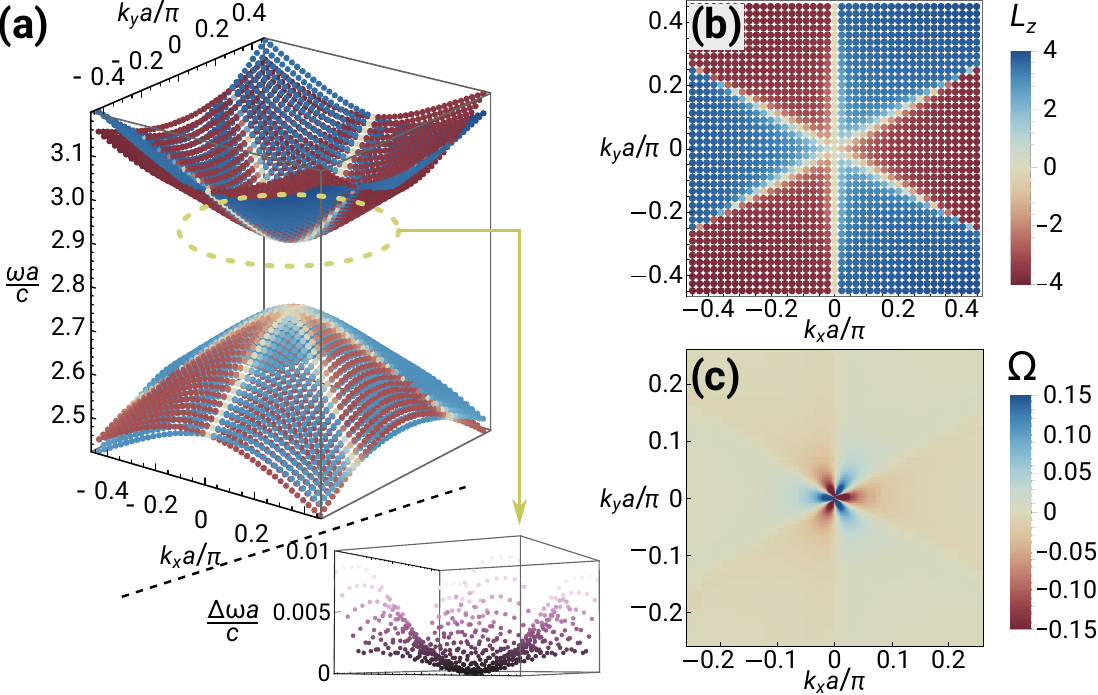}
\caption{\label{fig:Lz} (Color online) (a) Color-coded distribution of the pseudoangular momentum superimposed to the dispersion surface. The inset shows the frequency difference between the two upper bands, indicating nondegeneracy except at $\Gamma$. (b) shows the pseudoangular momentum of the fourth band ($\omega_3$). (c) Berry curvature corresponding to the lowest band ($\omega_1$). Results corresponding to the type-I lattice with $\phi=30^\circ$.
}
\end{figure}

This perturbation approach yields two 2-fold degenerate upper (lower) pseudospin bands, with isotropic circular cone-like dispersion.
However, this formulation is only valid in the neighborhood of the $\Gamma$ point and for small levels of distortion. In practice, the degeneracy only occurs at the $\Gamma$ point while the lattice symmetry only guarantees $C_{6v}$ symmetric dispersion. If the two pseudospin states are not degenerate, how are they distributed? It is certainly not possible for each band to represent a pseudospin, since it would imply that each band has a (pseudospin-resolved) Berry curvature that peaks at $\Gamma$, hence inconsistent with the premise of intact \textit{T}-symmetry.

To answer this question, we no longer use the 4-fold degenerate states as the basis, but instead extract pseudo parameters from the numerical solution. First, we calculate the pseudoangular momentum of each eigenstate in $\mathbf{k}$-space which allows tracking the pseudospin states. We define a pseudoangular momentum similarly to the expected value of $z$-angular momentum in the quantum mechanical case, $L_z = -i \braket{\mathbf u|\rho\partial_\theta|\mathbf u}$ \cite{Supp},
where $\mathbf{u}=(u_x,u_y,u_z)$ is the eigenstate displacement field, $\rho$ is the mass density, and $\theta$ is the angular coordinate of the cylindrical reference frame with the longitudinal axis parallel to $z$-axis and centered at the unit cell center (see Fig.~\ref{fig:lattice} (a)).

The resulting pseudoangular momentum is plotted in color on the corresponding dispersion curves over a square domain that is approximately $0.4|\Gamma K|$ in size (Fig.\ref{fig:Lz} (a)). Fig.\ref{fig:Lz} (b) show the pseudoangular momentum of the fourth band as an example. The lower inset in Fig.\ref{fig:Lz} (a) shows the frequency difference for the lower two states. This plot highlights that the two bands are detached everywhere except at the $\Gamma$ point.

This result can also be interpreted from a different perspective. One pseudospin state has a $C_{3v}$ symmetric dispersion with constant $L_z$ while the other pseudospin (that is the \textit{T}-counter part) has a dispersion surface that is $180^\circ$-reversed and displays opposite $L_z$. The coupling between the two-fold degenerate bands having opposite pseudospins results in band repulsion (along $\Gamma$-M directions) which leads to nondegeneracy except at $\Gamma$. It is found that the repulsion strength also depends on parameter $\phi$ \cite{Supp}. It is also worth noting that the value of $L_z$ is gauge dependent: unit cells with $\phi = \phi_0 + n (2\pi/3), \forall n\in \mathbb{N}$ all represent the same bulk lattice but $L_z$ varies depending on the number $n\mod 3$. This gauge transformation is equivalent to shifting the position of the longitudinal ($z$) axis of the cylindrical coordinates when calculating $L_z$. Nevertheless, with any fixed gauge choice, $L_z$ of inverted lattices (types-I/II obtained by applying the inversion operator about the unit cell center) always show opposite signs, therefore confirming that band inversion occurs during continuous morphing from one lattice to the other.

Fig.~\ref{fig:Lz} (c) shows the Berry curvature of the lowest band in a squared area ($\sim 0.25|\Gamma K|$) around the $\Gamma$ point \cite{Supp}. The Berry curvature concentrates near $\Gamma$ and shows a 6-lobe pattern as observed for $L_z$. A similar behavior is found for the remaining bands (i.e., both lower and upper bands) provided a change in signs. Clearly,  $\Omega(\mathbf{-k})=-\Omega(\mathbf{k})$ of each band is a consequence of \textit{T}-symmetry and it results in a vanishing accumulating Berry phase over the entire Brillouin zone. However, the pseudospin-resolved Berry phase is nonzero. We also note that, the Berry curvature of each band of type-II lattice, is identical to that of type-I but with opposite signs.
Numerical integration of the Berry curvature shows that the pseudospin Chern numbers are $C_{\uparrow/\downarrow}=\pm 1$, hence leading to nontrivial $\mathbb{Z}_2$ invariant $\frac{C_\uparrow-C_\downarrow}{2} \mbox{ mod } 2=1$ \cite{kaneCh1}. These topological invariants show that the bulk lattices have a non-trivial topological significance.

\begin{figure}[h]
\includegraphics[scale=0.7]{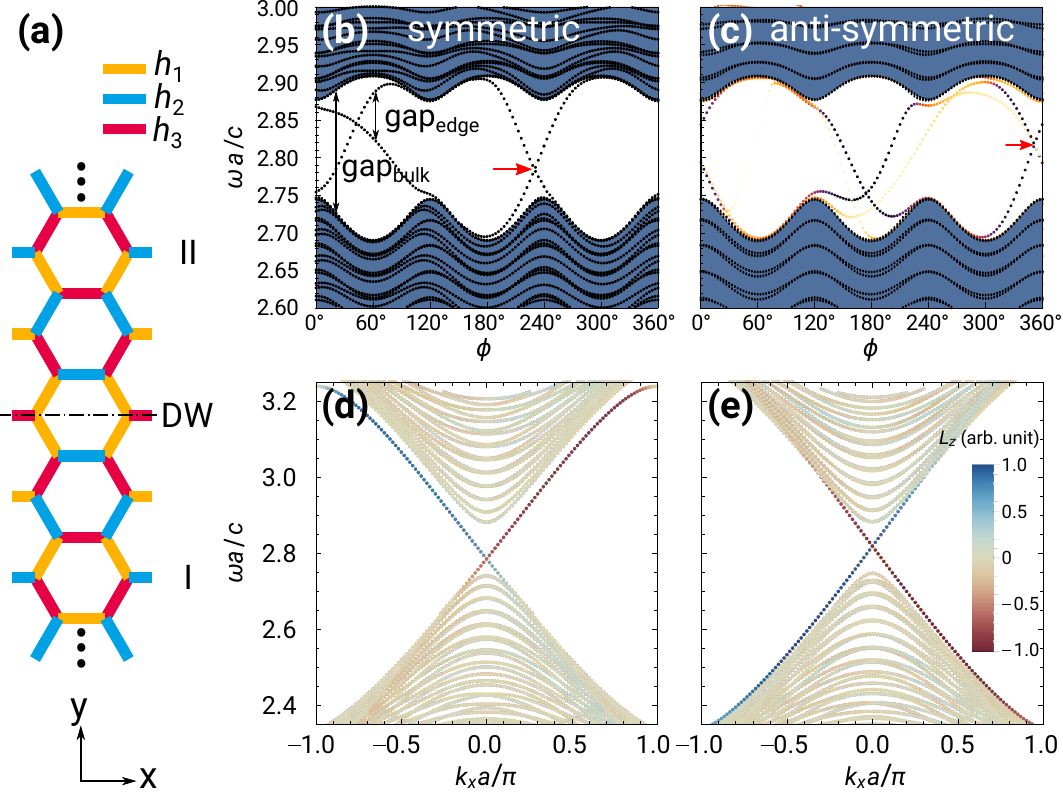}
\caption{\label{fig:gapless}
(Color online)  (a) Symmetric DW configuration. (b), (c) Eigenfrequencies of the supercell at $k_x=0$ versus $\phi$. Both symmetric and antisymmetric states are considered. (d), (e) The supercell dispersion corresponding to parameter indicated by the red arrow in (a) and (b), respectively.}
\end{figure}

Based on the information from both the asymptotic effective Hamiltonian and the first principle numerical calculations, topologically protected edge states should be expected at the interface between type-I and II lattices. Given that the two lattices are mirrored images to each other, a symmetric domain wall (DW) can be constructed, as shown in Fig.~\ref{fig:gapless} (a).

In electronic TIs, the crossing of the edge state at $\Gamma$ is protected by the degeneracy theorem of Kramers \cite{ReviewKaneTI}. In the acoustic case, it is no longer guaranteed given phonons are bosons. It follows that gaps are typically found in the edge states of most acoustic TIs based on zone-folding strategy \cite{MechanicalTI, AcousticTIAir, fold2017,acousticTI2017, zonefold2,kekuleprl2017, kekulenjp2018, zonefold2018, xia2018programmable, zhang2017topological, yves2017topological, vs, deng2017observation}.
In stark contrast with these cases, the generalized Kekul\'e distortion (Eq.\ref{eq:dist}) offers an additional degree of freedom that allows control on the dispersion of edge states. This degree of freedom is directly connected to the parameter $\phi$ that controls the breaking of \textit{P}-symmetry, which is intact when $\phi = n \pi /3$ and broken otherwise. The proper selection of this parameter yields gapless edge states by purposely exploiting the use of accidental degeneracies, which allow complete decoupling of counter-propagating (i.e., helical) edge states.
Note that topological materials relying solely on \textit{P}-breaking in honeycomb lattices belong to a different topological phase typically related to analogue QVHE \cite{ValleySonicBulk, ValleySonicEdge, pal2017edge, vila2017observation, myqv, QVEXP, diatom, yan2018chip, snowflake, ma2019valley}.

Fig.~\ref{fig:gapless} (b) and (c) plot the eigenfrequency for both symmetric and antisymmetric modes of the supercell at $k_x=0$ with $\phi$ $\in [0,2\pi]$. The blue shaded zone indicates the bulk bands, while the points in the bandgap are the edge states. The width of the gaps of both the bulk and the edge states are also shown. Recall that $\phi+2n\pi/3$ yields the same bulk lattice of $\phi$, so the bulk band pattern repeats every $2\pi/3$. However, the DW configuration changes with different $n$, so the edge states have no repeating pattern. The red arrows in Fig.~\ref{fig:gapless} (b) and (c) indicate examples where the gap of the edge states closes, while Fig.~\ref{fig:gapless} (d) and (e) plot the corresponding supercell dispersion. In these two cases we obtain helical edge states.
We also computed the pseudoangular momentum based on the cell next to the domain wall which is indicated in color to illustrated the pseudospin polarization.

However for the two lattices supporting edge states in between, it is possible to find a path in the parameter space $(\delta_1,\delta_2,\delta_3)$ linking the two configurations such that when the lattice evolves adiabatically along it, the bulk gap never closes, hence implying no topological phase transition. And both bulk lattices can have nontrivial $\mathbb{Z}_2=1$. This hold true for different (e.g., assymetric) types of DW assemblies \cite{Supp}.
These considerations indicate that, while these bulk lattices are seemingly indistinguishable from a topological perspective, they can still yield non-trivial edge states following translations perpendicular to the domain wall.
This observation is consistent with the fact that the existence of edge states on DWs between Kekul\'e lattices depends not only on the bulk properties of the two individual lattices but also on the same reference gauge choice. This result is a direct consequence of the gauge-dependence of both the $L_z$ calculation and the mapping to the BHZ model. Based on this idea, one can even devise helical edge states along an \textit{edge dislocation} of a Kekul\'e lattice
(see supplemental material \cite{Supp} for examples). Another related work recently reported the possibility to create DW between nontrivial phases controlled by different analogue quantum mechanisms \cite{miniaci2019valley}.

\begin{figure}[h]
\includegraphics[scale=0.38]{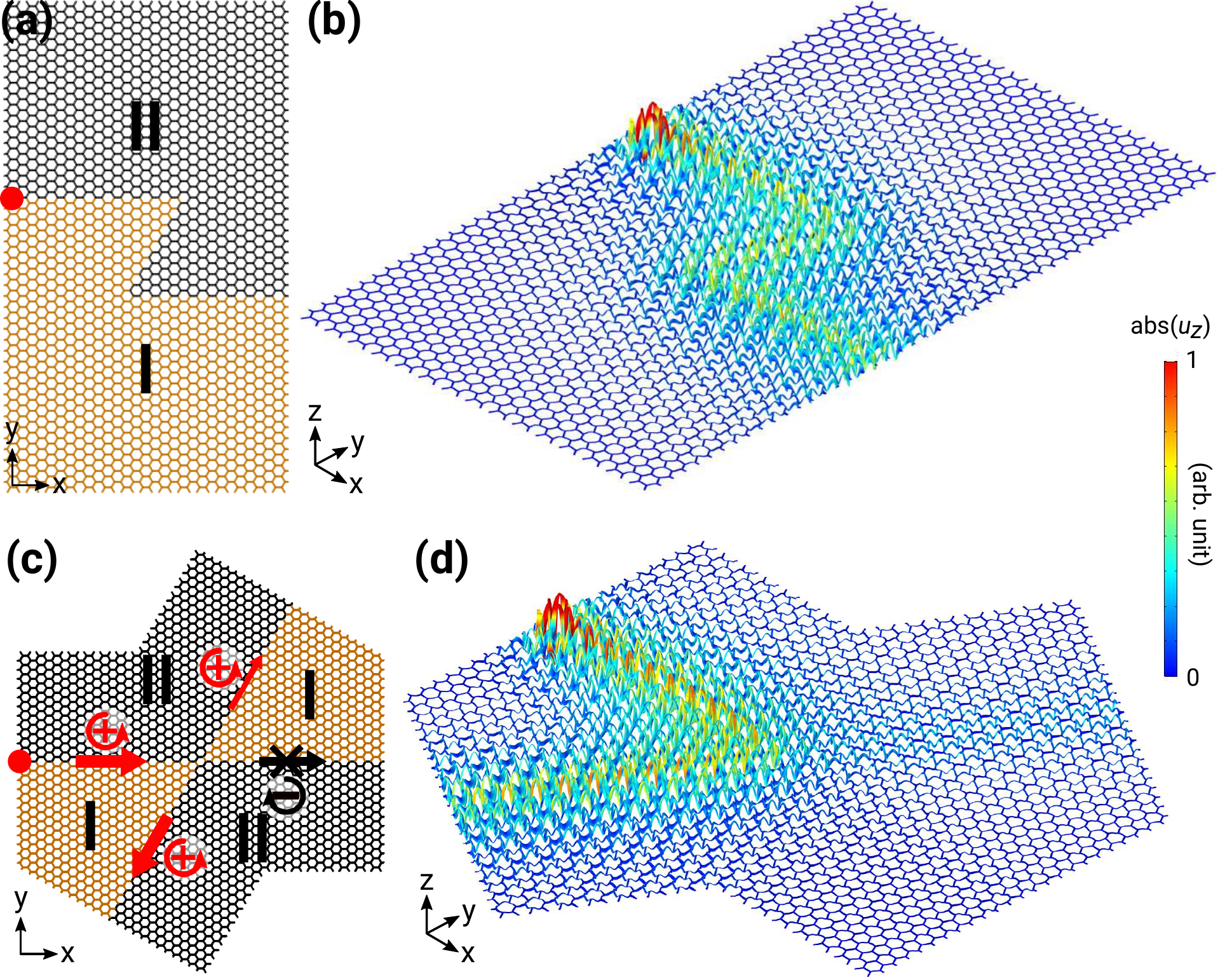}
\caption{\label{fig:ZX} (Color online) Steady-state harmonic response of excitation at DWs. (a) The configuration of a Z-shape DW. The red dot indicates the excitation and (b) the corresponding response. (c) The configuration of X-shape crossing DWs and (d) the corresponding response.}
\end{figure}

Finally we performed full-field numerical simulations to demonstrate the robustness of the edge states. Symmetric DW configuration with $\phi=229.9^\circ$ (see Fig.~\ref{fig:gapless} (b)) is selected in this case. Fig.~\ref{fig:ZX} (a) and (b) show both the configuration and the steady-state response of a Z-shaped DW under a harmonic excitation applied at one end (red dot). The edge state concentrates along the Z-path with uniform amplitude indicating no significant reflections at the corners. The second geometric configuration shows the case of an intersection between two DWs with excitation at one of the four terminals (Fig.\ref{fig:ZX} (c)).

In this case, the $+x$ propagating state is a pseudospin up. After the intersection, only the oblique branches support pseudospin-up out-going modes while the remaining wall along the $+x$ axis is associated with pseudospin down. 
The result is that the edge state continues propagating only along the two oblique walls.
The different amplitude of the mode along the two DW branches is caused by the different configuration of the walls. The lower branch has the same configuration as the input one, while the upper one is equivalent to a configuration with $ \phi -2 \pi /3$ (see Fig. \ref{fig:gapless} (a)) which is not optimized to supports gapless edge states at the selected frequency. This is an interesting feature of this DW and can be used to create complex patterns of propagation.

In summary, we proposed a continuous elastic phononic structure composed of slender beams and exploiting Kekul\'e distortion to achieve an elastic analogue of the spin Hall effect for flexural waves.
We derived the effective Hamiltonian based on conventional perturbation analysis and pseudospin bases and showed that it maps to the BHZ model for QSH materials.
Furthermore, first-principles calculations revealed the finer structure of the Kekul\'e lattices that further the understanding of these material phases beyond the low-energy Hamiltonian approximation available in the literature. The finer structure includes a peculiar 6-lobe alternating pattern of both the pseudospin texture and the Berry curvature.
We also investigated the unique gauge dependence of the topological states and its connection to the occurrence of helical edge modes. A remarkable consequence of this finding is the possible existence of topological edge states on edge dislocations of a lattice.
We show, also for the first time, that the gapless edge states in an elastic Kekulé lattice (associated with a degeneracy at the $\Gamma$ point) can always be achieved by combining lattice symmetry and accidental degeneracies.
Numerical simulations confirmed that the counter-propagating pseudospin polarized edge states are decoupled and robust to disorder.

\begin{acknowledgments}
\section{Acknowledgments}
The authors gratefully acknowledge the financial support of the National Science Foundation under grant MOMS \#1761423.                                                                                                                                                                                                                                                                                                                                                                                                                                                                                                                                                                                                                                                                                                                                                                                                                                                                                                                                                                                                                                                                                                                                                                                                                                                                                                                                                                                                                                                                                                                                                                                                                                                                                                                                                                                                                                                                                                                                                                                                                                                                                                                                                                                                                                                                                                                                                                                                                                                                                                                                                                                                                                                                                                                                                                                                                                                                                                                                                                                                                                                                                                                                                                                                                                                                                                                                                                                                                                                                                                                                                                                                                                                                                                                                                                                                                                                                                                                                                                                                                                                                                                                                        
\end{acknowledgments}


\bibliography{REF_v3}

\end{document}



\title{Supplemental material: Topological properties and helical edge states in elastic phononic waveguides with Kekul\'e distortion}

\author{Ting-Wei Liu}
\author{Fabio Semperlotti}
 \email{fsemperl@purdue.edu}
\affiliation{Ray W. Herrick Laboratories, School of Mechanical Engineering, Purdue University, West Lafayette, Indiana 47907, USA.}

\date{\today}
\maketitle
\section{Effective Hamiltonian}

\begin{figure}[h]
\includegraphics[scale=0.8]{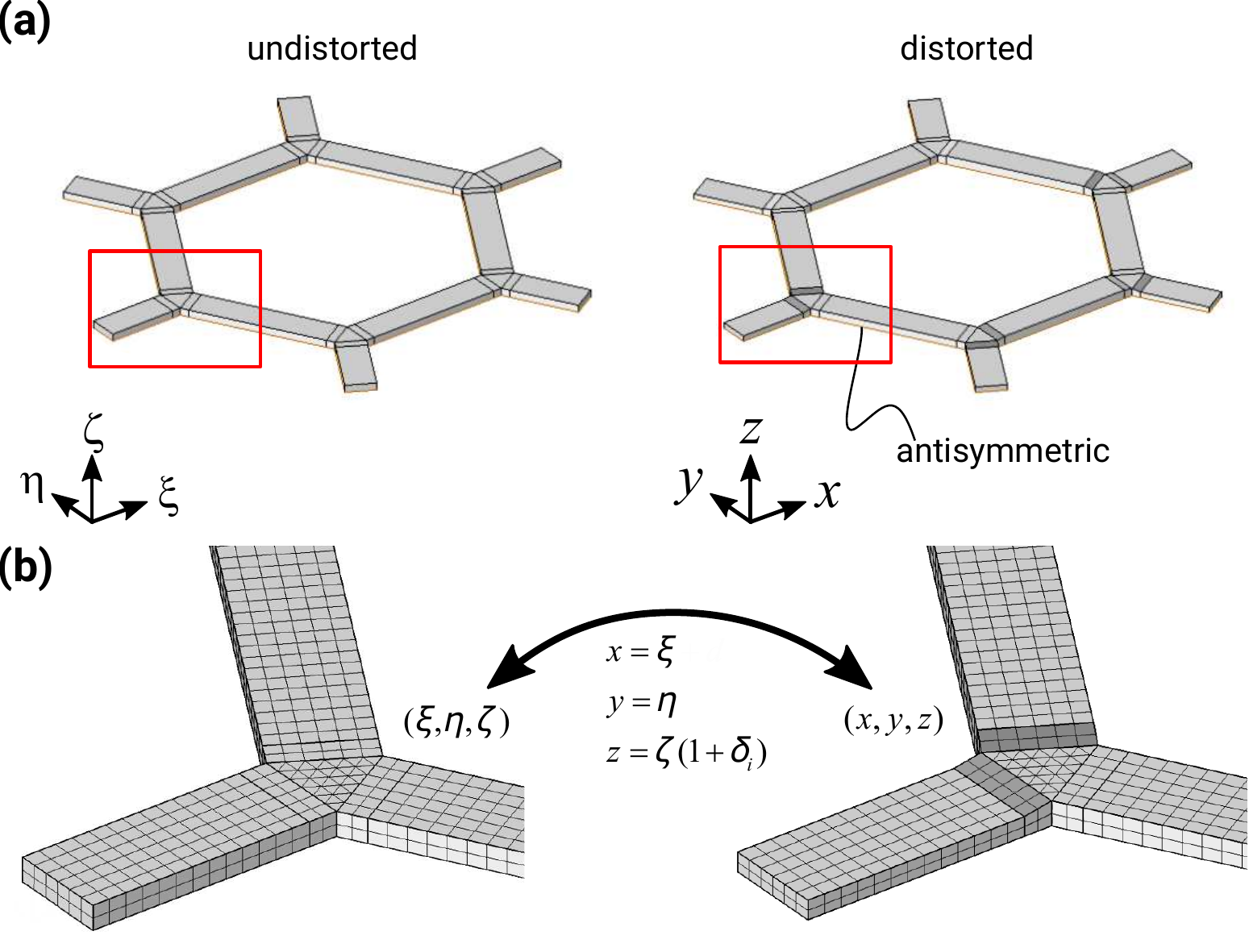}
\caption{\label{fig:map} (Color online) (a) Schematic showing the undistorted and the distorted configurations and the corresponding reference frames and mapping operations. (b) Detailed view of the finite element models used in for the mapping procedure.
}
\end{figure}

The numerical calculation presented in this study follow from a direct solution of the elastodynamic Navier's equations for linearly elastic isotropic solids,
\begin{equation}
(\lambda+2\mu)\nabla \nabla \cdot \mathbf{u}-\mu \nabla^2 \mathbf{u}=\rho \frac{\partial^2\mathbf{u}}{\partial t^2},
\end{equation}
where $\lambda$ and $\mu$ are the elastic Lam\'e constants, $\mathbf{u}$ the displacement vector, and $\rho$ the mass density. These equations are solved using the commercial finite element package COMSOL Multiphysics. The Bloch states of a given wavevector $(k_x,k_y)$ are obtained by calculating the eigenmodes of the unit cell with Bloch periodic boundary conditions. In this study, we focus on flexural waves in the reticular plates, that is antisymmetric Lamb modes. Therefore, the finite element model can be reduced to the half plate with antisymmetric boundary conditions applied on the mid-plane ($xy$-plane), as shown in Fig.~\ref{fig:map}.

To derive the effective Hamiltonian of the lattice with Kekul\'e distortion, we assume that the eigenstates can be expressed as an expansion of the four degenerate states of the undistorted lattice (in the following referred to as the \textit{undistorted basis representation}). Also the basis is arranged in the pseudo spin fashion, by linear combination of symmetry adapted eigenstates, as depicted in the online supplemental video \cite{SuppV}.

To find the expansion coefficients, we evaluate the inner product integral of the undistorted basis $\ket{b_i}$ and the distorted eigenstates $\ket{v_j}$ which have been both previously normalized such that $\braket{b|\rho|b}=\braket{v|\rho|v}=1$, where $\rho$ is the mass density. Note that $\ket{b_i}$ and $\ket{v_j}$ are defined in different spaces that describe lattices with different shapes. In order to evaluate the integral involving two functions defined in two different domains, we first need to project the states onto the same space. We adopt the following mapping strategy. Consider a (finite element) model for each geometry, as shown in Fig.~\ref{fig:map} (a). A piecewise homeomorphic mapping relation exists between these two models, as shown in Fig.~\ref{fig:map} (b). The deformation gradient of the mapping for each beam can be written as
\begin{equation}\label{F}
\mathbf{F}=\frac{\partial \mathbf{x}}{\partial \bm{\xi}}
=\left (
\begin{array}{ccc}
 1& 0& 0\\
0&1&0\\
 0&0&1+\delta_i\\
\end{array}
\right ),
\end{equation}
where $\mathbf{x}=(x,y,z)$, $\bm{\xi}=(\xi,\eta,\zeta)$ as depicted in Fig.~\ref{fig:map}, and $i = 1,2,3$ stands for the three beams.

Once the eigenstates defined on the distorted geometry are mapped back onto the undistorted geometry, one can take the inner product of both sets in the undistorted basis and determine the four corresponding components.
Note that in the distorted lattice, the stored energy per unit length in the beam is proportional to the deformation amplitude and the beam height. When mapping the (distorted) eigenstates back to the undistorted lattice with uniform beam heights, the energy distribution on the three beams are scaled unevenly (with factors reciprocal to the beams' height) and does NOT accurately describe the true energy distribution. Therefore, an additional scaling factor accounting for each beam height is included.
The distorted eigenstate $\ket{v_j}$ (a continuous spatial vector field with infinite degrees of freedom) under undistorted basis representation, is then simplified to an $1\times 4$ vector $\mathbf{v}_j$. 
Explicitly, for the $j^{th}$ distorted eigenstate, its $k^{th}$ component under undistorted basis representation is computed by
\begin{equation}\label{vi}
\mathbf{v}_j(k)=\braket{b_k|v_j}_{\textrm{undistorted space}}\\
=\sum_i^3 (1+\delta_i)\int_{\textrm{beam}_i} \rho \mathbf{u}_{b_k}^\dag \mathbf{u}_{v_j} dxdydz + \int_{\textrm{prisms}} \rho \mathbf{u}_{b_k}^\dag\mathbf{u}_{v_j} dxdydz,
\end{equation}
where $\mathbf{u}$ represents the eigenstate's displacement vector field.
Note that, as for any perturbation method, the expansion is accurate only for small distortion strengths. In our analyses, we apply a distortion strength $\delta=25\%\times h_0$ which is considered acceptable in view of the Parseval's identity (the evaluated norm of each state under undistorted basis representation is still close to 1, $\lVert\mathbf{v}_j\rVert^2=1\pm 17\%$).
Note that we checked that $\braket{v_i|\rho|v_j}=\delta_{ij}$ is still numerically accurate (less than 1\% error) when performing the inner product on the undistorted cell with proper scaling factors (accounting for the beam thickness). It follows that the 17\% error does not come from the mapping operation but instead it is due to the distortion strength (in the sense that the undistorted basis cannot accurately resolve the distorted eigenstates). Nevertheless, this perturbation approach still provides some useful insights.

The eigenvalue problem of the Bloch state is given by $\hat H\ket{v_j}=\omega_j^2\ket{v_j}$ ($\hat H$ is the Hamiltonian operator, and $\ket{v_j}$ is the distorted eigenstate in the $xyz$-representation). The approximate form of $\hat H$ in the undistorted basis representation is given by a $4\times 4$ effective Hamiltonian $H_{mn}$

\begin{equation}\label{eq:Hmn1}
\mysq{H_{mn}}
\left \{
\begin{array}{c}
 \\
\mathbf{v}_j \\
 \\
\end{array}
\right \}
=
\omega_j^2
\left \{
\begin{array}{c}
 \\
\mathbf{v}_j \\
 \\
\end{array}
\right \}.
\end{equation}
In extended form the Hamiltonian can be written as
\begin{equation}\label{eq:Hmn2}
\mysq{H_{mn}}
=
\left(
\begin{array}{c|c|c|c}
 &  &  &  \\
\mathbf{v}_1 & \mathbf{v}_2 & \mathbf{v}_3 & \mathbf{v}_4 \\
 &  &  &  \\
\end{array}
\right)
\mydiag{\omega_j^2}
\left(
\begin{array}{c|c|c|c}
 &  &  &  \\
\mathbf{v}_1 & \mathbf{v}_2 & \mathbf{v}_3 & \mathbf{v}_4 \\
 &  &  &  \\
\end{array}
\right)^{-1}.
\end{equation}
Given that $\mathbf{v}_j$ and $\omega_j$ are known, we can evaluate the Hamiltonian at $\Gamma$ as
\begin{equation}
H|_\Gamma =\left(
\begin{array}{cc|cc}
\fbox{$-8.52$}-0.00i  &    +0.00-0.01i      &    +0.01+0.00i       &  -0.00+3.20i\\ 
+0.00+0.01i           &\fbox{$+8.52$}-0.00i &    -0.00+3.49i       &  +0.01-0.00i \\ \hline
+0.01-0.00i           &    -0.00-3.20i      &\fbox{$-8.52$}+0.00i  &  +0.00+0.01i\\
-0.00-3.49i           &    +0.01+0.00i      &    +0.00-0.01i     &\fbox{$+8.52$}+0.00i \\
\end{array}\right)\times 10^8
\end{equation}

Recall that in the effective Hamiltonian
$ h(\mathbf{k}) = (C-Dk^2)\sigma_0 + \EuScript{A}(k_x\sigma_x-k_y\sigma_y) + (M+Bk^2)\sigma_z$
the $\sigma_0$ components stand for an overall shifting of the dispersion curves and have no topological significance. $C$ is chosen to be 0 by shifting the frequency datum to the degenerate frequency of the undistorted lattice. $D$ is calculated as $7.02\times 10^5$. Later in the procedure, $D$ is filtered out from the second order derivatives of the Hamiltonian (Eq.~\ref{eq:d2Hdkx2} and \ref{eq:d2Hdky2}). This step is performed in order to improve the clarity of the matrix structure.  Neglecting the smaller imaginary antidiagonal elements, the effective Hamiltonian becomes a diagonal matrix with $-8.52$ $\sigma_z$ components in both $2\times 2$ block diagonal submatrices. This $\sigma_z$ component at $\Gamma$ is the massive term in the Dirac Hamiltonian and it indicates the existence of coupling between the $g$- and $h$-like states as well as the gap width at $\Gamma$. It follows that $M=-8.52$.

The dependence of the Hamiltonian from $k$ and $k^2$ in a small neighborhood around the $\Gamma$ point can be found by Taylor series expansion by taking 5 points on the $k_x$ and $k_y$ axes with step $\Delta k =0.002 \pi/a$ and finite difference approximation of the derivatives.

\begin{equation}
\left.\frac{\partial H}{\partial k_x}\right|_\Gamma =\left(
\begin{array}{cc|cc}
-0.00+0.00i           &\fbox{$-1.14+0.66i$} &    +0.01+0.01i       &  +0.00-0.00i\\ 
\fbox{$-1.29-0.74i$}  &    +0.00-0.00i      &    +0.00-0.00i       &  -0.01+0.01i\\ \hline
-0.01+0.01i           &    -0.00-0.00i      &    +0.00+0.00i       &\fbox{$+1.13+0.65i$}\\
-0.00-0.00i           &    +0.01+0.01i      &\fbox{$+1.30-0.75i$}  &  -0.00-0.00i\\
\end{array}\right)\times 10^8
\end{equation}

\begin{equation}
\left.\frac{\partial H}{\partial k_y}\right|_\Gamma =\left(
\begin{array}{cc|cc}
+0.00+0.00i           &\fbox{$-0.66-1.14i$} &    -0.01+0.01i       &  +0.00+0.00i\\ 
\fbox{$-0.74+1.29i$}  &    -0.00-0.00i      &    +0.00+0.00i       &  +0.01+0.01i\\ \hline
+0.01+0.01i           &    -0.00+0.00i      &    -0.00+0.00i       &\fbox{$+0.65-1.13i$}\\
-0.00+0.00i           &    -0.01+0.01i      &\fbox{$+0.75+1.30i$}  &  +0.00-0.00i\\
\end{array}\right)\times 10^8
\end{equation}

\begin{equation}\label{eq:d2Hdkx2}
\left.\frac{\partial^2H}{\partial k_x^2}\right|_\Gamma =
\left(
\begin{array}{cc|cc}
\fbox{$-2.24$}-0.00i  &    -0.43+0.24i      &    +0.09-0.16i       &  +0.00-0.12i\\ 
-0.39-0.22i           &\fbox{$+2.24$}+0.00i &    -0.00+0.25i       &  -0.09+0.16i\\ \hline
-0.09+0.16i           &    +0.00+0.12i      &\fbox{$-2.24$}+0.00i  &  -0.43-0.24i\\
-0.00-0.25i           &    +0.09+0.16i      &    -0.40+0.23i       &\fbox{$+2.24$}-0.00i\\
\end{array}\right)\times 10^6
\end{equation}

\begin{equation}\label{eq:d2Hdky2}
\left.\frac{\partial^2H}{\partial k_y^2}\right|_\Gamma =
\left(
\begin{array}{cc|cc}
\fbox{$-2.24$}+0.00i  &    +0.43-0.25i      &    +0.09+0.16i       &  -0.00-0.12i\\ 
+0.39+0.23i           &\fbox{$+2.24$}-0.00i &    -0.00+0.25i       &  +0.09-0.16i\\ \hline
+0.09-0.16i           &    -0.00+0.12i      &\fbox{$-2.24$}-0.00i  &  +0.43+0.25i\\
-0.00-0.25i           &    +0.09+0.16i      &    +0.40-0.24i     &\fbox{$+2.24$}+0.00i\\
\end{array}\right)\times 10^6
\end{equation}

The first order derivative with respect to $k_x$ ($k_y$) corresponds to the $\sigma_x$ ($\sigma_y$) component, i.e., the matrix of coefficients $\EuScript A$. It also indicates the group velocity of the Dirac bands (i.e. before lifting the degeneracy). The second order derivative with respect to $k$ shows the $\sigma_z$ pattern, which maps to the matrix of coefficients $B$ and indicates the curvature of the gap opening at $\Gamma$.

It is known that whether or not the gap coefficient ($M$) and the curvature coefficient ($B$) have the same sign, indicates the topological character of the lattice. In our case, we find $B/M=2.63\times 10^{-3}>0$ indicating that this is a nontrivial lattice \cite{bhz}. We note, however, that given the same bulk lattice with a different gauge choice, the Hamiltonian is not unique. For example, while the following substitution $(\delta_1,\delta_2,\delta_3)\rightarrow (\delta_2,\delta_3,\delta_1)$ (or equivalently $(\delta_1,\delta_2,\delta_3)\rightarrow (\delta_3,\delta_1,\delta_2)$) results in a new unit cell geometry, the new cell is simply a translated version of the original bulk lattice (or, equivalently, of the reference frame). At the same time, the corresponding Hamiltonians are different and the values of $B/M$ are $-2.16\times 10^{-3}<0$ and $-2.57\times 10^{-2}<0$, respectively. This observation (different signs from the previous obtained value $B/M=2.63\times 10^{-3}>0$) implies that the topological character based on mapping to the BHZ model  depends on the gauge choice.

\section{Additional results on the dispersion surfaces and the pseudoangular momenta}

The pseudoangular momentum is computed using the following equation,
\begin{equation}\label{eq:Lz}
   L_z =
   -i \braket{\mathbf u|\rho\partial_\theta|\mathbf u},
\end{equation}
where the eigenstate $\ket{\mathbf{u}}$ are obtained numerically from the finite element model. This quantity is conceptually similar to the expected value of $z$-angular momentum in the quantum mechanical case.
In quantum mechanics, the normalization of eigenstates is given by a unitary total probability $\braket{\psi^{(m)}|\psi^{(n)}}=\delta_{mn}$. For classical waves there is no restriction on amplitude. Here, we choose the following normalization $\braket{\mathbf u^{(m)}|\rho|\mathbf u^{(n)}}=\delta_{mn}$ such that the total mechanical energy of each state is normalized to $\frac{1}{2}\omega^2$ (The kinetic energy amplitude density at each point in the solid is $\frac{1}{2} \rho\omega^2\mathbf{u}^\dagger \mathbf{u}$ and it also gives the total mechanical energy density). Note that in the present case, this $L_z$ quantity does not refer to a rotating mass while, instead, it measures the strength of the mechanical energy circulation.

\begin{figure}[h]
\includegraphics[scale=0.6]{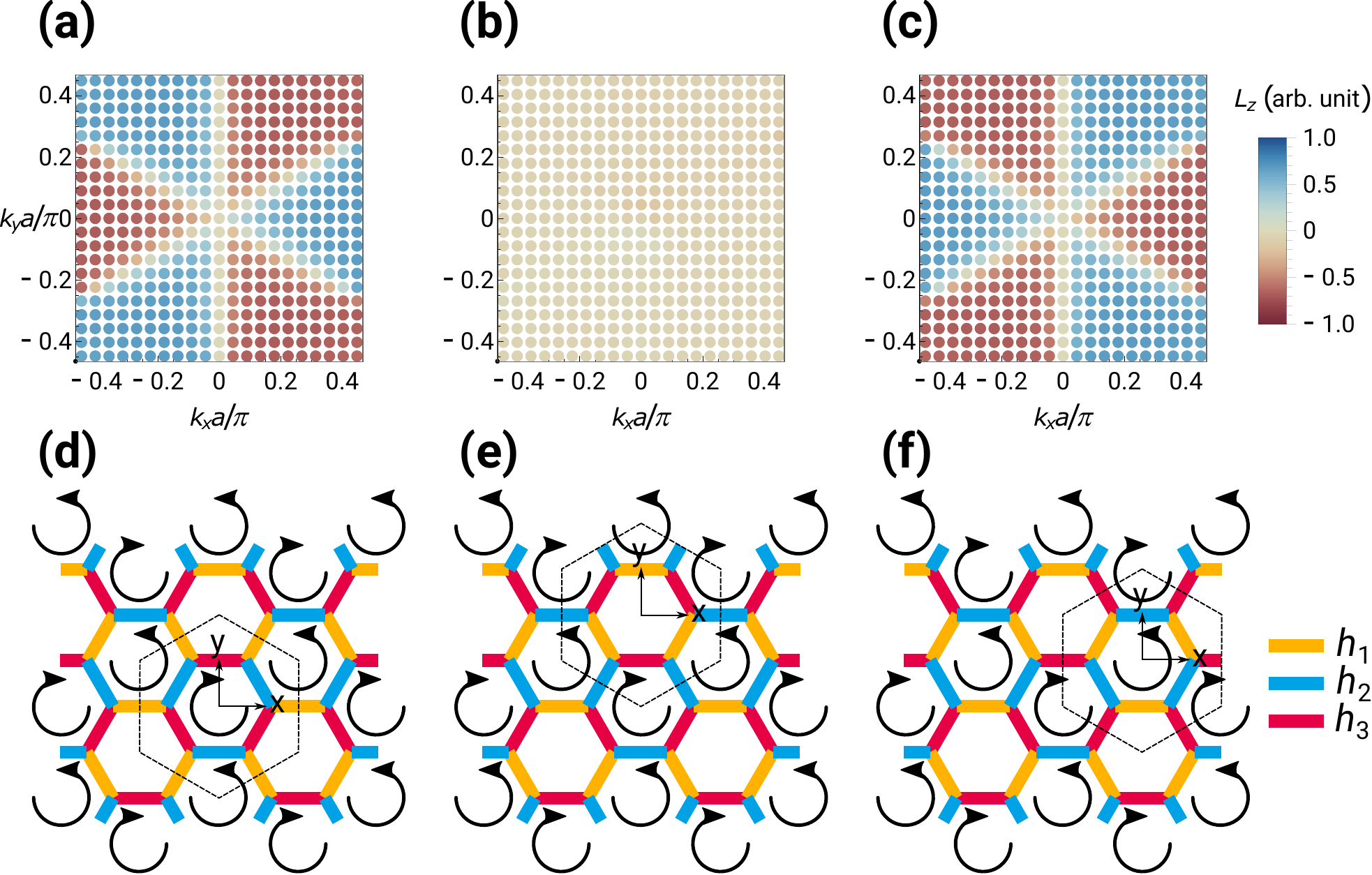}
\caption{\label{fig:Lz}  (a-c), The pseudoangular momenta of the fourth band for the case $\phi = 60^\circ$ and for different unit cell choices (dashed hexagons in (d-f), respectively) of the same bulk lattice. (d-f) The schematics of the lattice and the choice of the unit cell. Circular arrows indicate the local mechanical energy flow.}
\end{figure}

In Fig.~\ref{fig:Lz} (a-c), we show the pseudoangular momenta of the fourth band for the case $\phi = 60^\circ$ and for different unit cell choices of the same bulk lattice. The choices of the unit cell are indicated by dashed hexagons in (d-f) of the same figure. Results show that the pseudoangular momenta can be different (with opposite sign, or zero) given shifted reference frames. Such difference is analogous to that observed in the calculation of classical angular momenta with respect to different axes.

The (counter-)clockwise arrows in Fig.~\ref{fig:Lz} indicate the local mechanical energy flow at $k_x=0.4\pi/a, k_y=0$. In this specific example ($\phi=60^\circ$), the lattice has higher symmetry ($C_{6v}$) than the general case ($C_{3v}$). The two counter-rotating energy flows have equal strength, hence resulting in zero circulation on some hexagons.

\begin{figure}[h]
\includegraphics[scale=0.5]{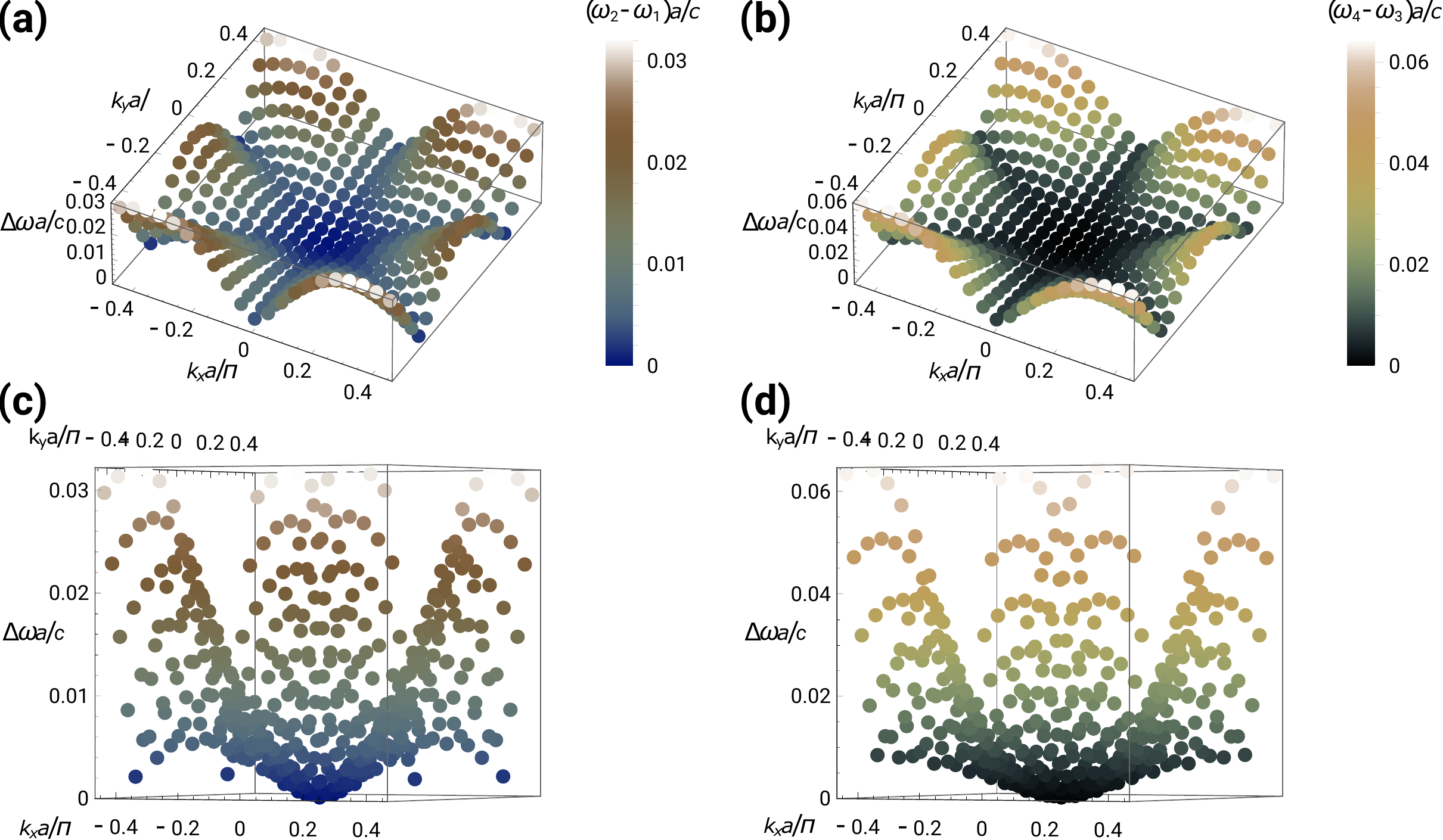}
\caption{\label{fig:diff} (a,c) The frequency difference of the two bands of the lower cone bands ($\omega_2-\omega_1$). (b,d) The frequency difference of the two bands of the upper cone bands ($\omega_4-\omega_3$). Both show that the frequency dispersion surfaces are degenerate only at a single point $\Gamma$.}
\end{figure}

Fig.~\ref{fig:diff} (a,c) and (b,d) show the frequency difference between the two bands of either the lower cone ($\omega_2-\omega_1$) or the upper cone ($\omega_4-\omega_3$), respectively. Both show that the frequency dispersion surfaces are degenerate only at a single point $\Gamma$.

\section{Calculation of the Berry curvature}

The Berry connection $\mathbf{A}^{(n)}(\mathbf{k}) = i\braket{\mathbf{u}^{(n)}(\mathbf{k})|\nabla_\mathbf{k}\mathbf{u}^{(n)}(\mathbf{k})}$ can also be calculated directly from first principles without any approximation. Given that the eigenstates are nondegenerate, Berry curvature can be explicitly calculated as
$\Omega_{xy}^{(n)}=\partial_{k^x} A_y^{(n)} - \partial_{k^y} A_x^{(n)}$.
In this expression, $\ket{u}$ is the periodic part extracted from the Bloch state $\ket{v}=\ket{u}e^{i\mathbf{k\cdot r}}$. The calculation of the normalized eigenstate is still associated with a phase ambiguity $\ket{v}e^{i\theta}$.
Theoretically, such ambiguity should not affect the Berry curvature (equivalently to any gauge field).  However, if the phase varies randomly as a non-differentiable function of  $(k_x,k_y)$, it may still cause issues when evaluating the derivative of an eigenstate in $k$-space, such as in the case of the Berry connection expression calculation. In our case, we use the finite element package COMSOL to numerically extract the eigenstates. It is not guaranteed that the phase ambiguity is smooth as the solver sweep through different $(k_x,k_y)$ points. To eliminate this apparent ambiguity, we use a reference point for the phase. All eigenstates are subtracted by the phase of a particular point (specifically, vertex \# 1 of the FE model), such that the phase at that point is always 0. In this way the resulting phase ambiguity is at least smooth in $k$-space.

\section{Edge states supported by different DW configurations}
\begin{figure}[h]
\includegraphics[scale=1]{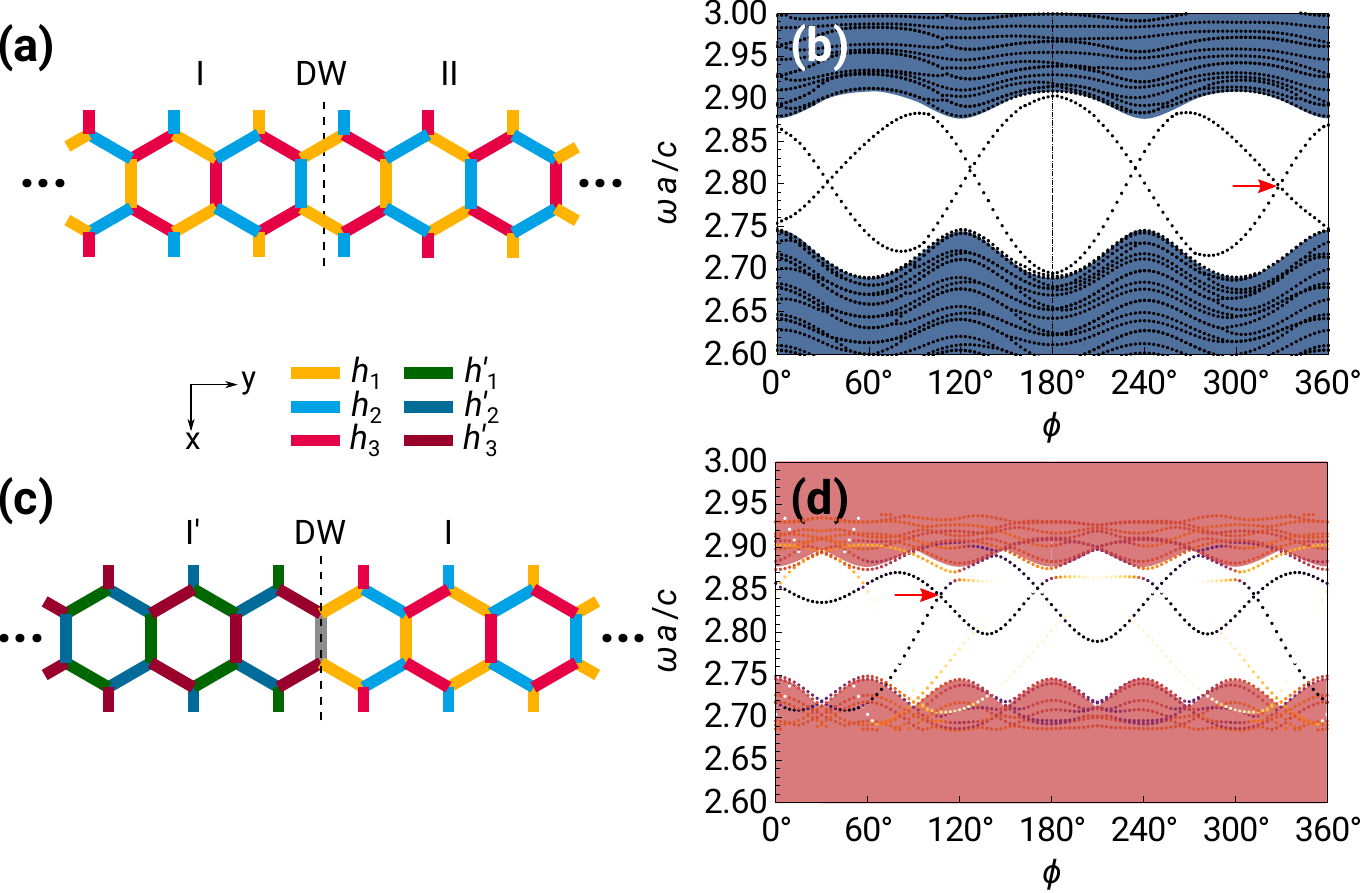}
\caption{\label{fig:sweepasymm} Schematic illustrations and the $k_x=0$ eigenfrequencies vs. $\phi$ of the supercell, for two types of asymmetric DW configurations. (a, b) Lattices I and II sharing common beams. (c, d) Lattices with distortions negative to each other.}
\end{figure}

In the main text, we have shown the case of edge states supported by a symmetric type of DW. The DW can also be built following an asymmetric configuration. We anticipate that in this case we cannot expect the edge states to be either symmetric nor antisymmetric with respect to the DW. Fig.~\ref{fig:sweepasymm} (a) shows a particular example where lattice I and II share common beams of thickness $h_1$ (yellow) at the DW interface. Fig.~\ref{fig:sweepasymm} (b) shows the corresponding supercell eigenfrequency at $k_x=0$ as a function of the parameter $\phi$.
Not surprisingly, the bulk band (blue shaded area) is identical to those of Fig.~6 in the main text. We find a period of $2 \pi/3$ in $\phi$ in the bulk bands since the bulk lattice remains the same pattern even when $\phi$ is substituted by $\phi+2\pi/3$. However, the edge state frequency does not possess that periodicity since the local pattern at the interface indeed changes under the same substitution $\phi \rightarrow \phi+2\pi/3$. Crossing of black dotted curves in the bulk gap indicates the parameter $\phi$ such that gapless edge state appears.

Previous studies \cite{kekuleprl2017,kekulenjp2018} also proposed another possible way to construct topologically protected states in Kekul\'e lattices by changing the \textit{sign} of the distortion (Fig.\ref{fig:sweepasymm} (c))  rather than using different permutations of $\delta_{1,2,3}$ as explored here above. For completeness, we investigated this design also for our continuous elastic system and report the cases of edge states. The height of the distorted pattern can be expressed as:
\begin{equation}
\begin{cases}
h_i = h_0 + \delta_i &\text{in lattice I}\\
h_i' = h_0 - \delta_i &\text{in lattice I}'
\end{cases}.
\end{equation}
Since $\delta_i = \delta \cos\big(\phi + (i-1)\frac{2\pi}{3}\big)$, the two can be considered as lattices whose respective parameter $\phi$ differs by $\pi$ from each other. As a result, the bulk band (pink shaded area shown in Fig.~\ref{fig:sweepasymm} (d)) consists of two copies of the bulk bands in Fig.~\ref{fig:sweepasymm} (b) displaced by $\pi$ and perfectly overlapped. The same situation can also result in the two lattices having different band gap ranges at a certain $\phi$. An effect of somewhat lower importance is that the penetration depth of the edge states is different in the two lattices, hence forming an asymmetric displacement amplitude distribution (See Fig.~\ref{fig:uneven}).

\begin{figure}[h]
\includegraphics[scale=1.35]{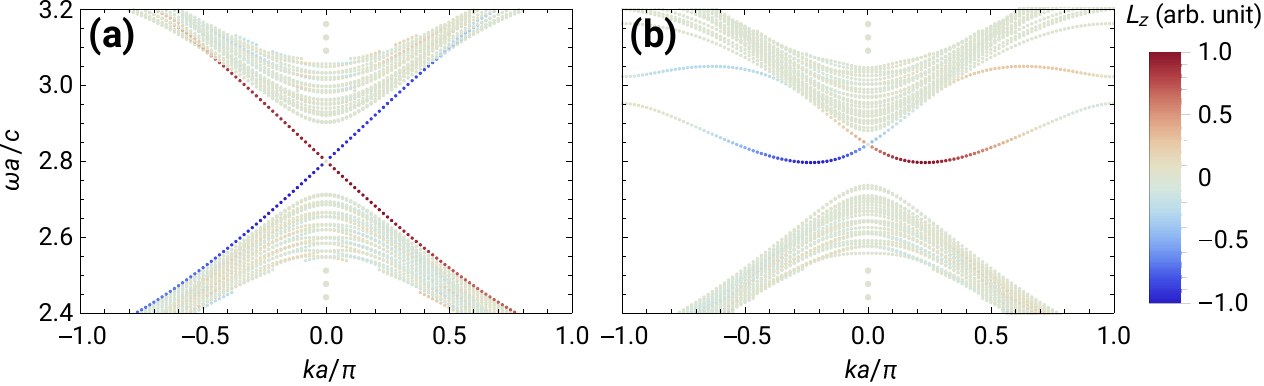}
\caption{\label{fig:asymmgapless} The DW supercell dispersion for the two asymmetric DW cases. (a) The first case with parameter $\phi=327.8^\circ$, as indicated by the red arrows in Fig.\ref{fig:sweepasymm} (b). (b) The second case with $\phi=107.0^\circ$, as indicated by the red arrows in Fig.\ref{fig:sweepasymm} (d).
}
\end{figure}

\begin{figure}[h]
\includegraphics[scale=1.6]{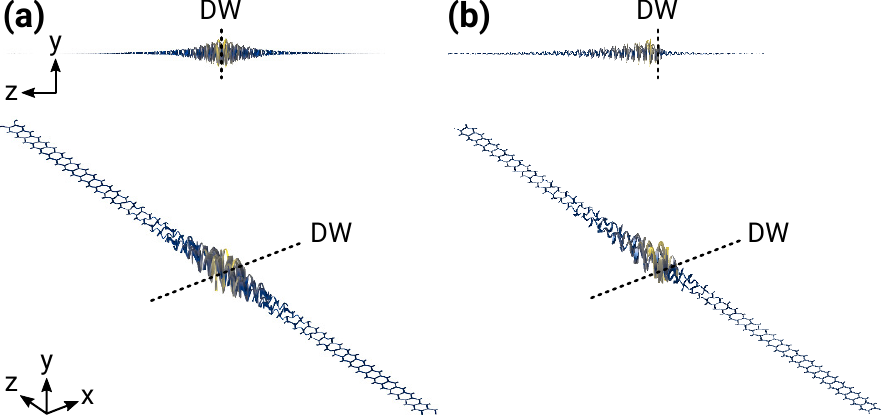}
\caption{\label{fig:uneven} The displacement fields of the edge states for the two cases corresponding to Fig.~\ref{fig:asymmgapless} (a) and (b), respectively. (a) The two lattices have the same bulk band gap, therefore the same evanescent behavior in both sides. (b) The two lattices have different gap sizes so that the edge states exhibit different penetration in the bulk.}
\end{figure}

Fig.~\ref{fig:asymmgapless} (a) and (b) show the DW supercell dispersion for the above two asymmetric DW cases, with parameter $\phi=327.8^\circ$ and  $\phi=107.0^\circ$, as indicated by the red arrows in Fig.\ref{fig:sweepasymm} (b) and (d), respectively.

Fig.~\ref{fig:uneven} (a) and (b) show the displacement fields of the edge states, for the two cases corresponding to Fig.~\ref{fig:asymmgapless} (a) and (b), respectively. In Fig.~\ref{fig:asymmgapless} (a), although the edge state is supported by an asymmetric DW the two lattices have the same bulk band gap, hence they exhibit the same evanescent behavior on both sides. In (b), the two lattices have different gap sizes hence resulting in different penetration of the edge state in the corresponding lattice.

Note that inverting the sign of the distortion is formally equivalent to offsetting $\phi$ by $\pi$. Such two lattice configurations can continuously evolve from one to the other while maintaining fixed distortion strength $\delta$ and varying $\phi$ continuously. During this process the bulk gap never closes, hence implying no topological phase transition. A similar result can be reached for the case of two lattices having inverted order of distortion: one can fix $\delta_1\neq 0$ (ensuring the bandgap) and gradually change the values of $\delta_2$ and $\delta_3$ until they completely exchange values (i.e.  $\delta_2\to\delta_3$ and $\delta_3\to\delta_2$). One will find that, for the two inverted lattices supporting the edge state on their common interface, both bulk lattices have nontrivial $\mathbb{Z}_2=1$.

Another interesting observation can be drawn at this point. In the neighborhood of $\phi=120^\circ$ and $240^\circ$  we have gapless edge states. Similarly, edge states appear around $\phi=180^\circ, 360^\circ$ in Fig.~6 (b) in the main text.
When $\phi=n\pi/3$, it is either $h_1 = h_2$, $h_2 = h_3$, or $h_3 = h_1$, which means practically that type-I and II lattices (with inverted the cyclic permutation of $h_{1,2,3}$) have the same pattern and differ only by a relative translation.
It means that, when $\phi=n\pi/3$, the DW interface connects two pieces of the same type of bulk lattice with a finite relative translation between them. This result suggests that, under proper conditions, topological gapless edge states can even exist along an \textit{edge dislocation} of a single type of lattice.

This result also further substantiates the argument proposed in the main text according to which we cannot predict the existence of edge states on a DW, simply by assessing the values of the $\mathbb{Z}_2$ invariant of the two connected lattices. As already mentioned, the DW configuration varies with the substitution $\phi\rightarrow\phi+2n\pi/3$ (equivalent to a relative rigid translation of the two sub-domains) while the bulk patterns remain the same. The corresponding spin Chern numbers can also vary given they are gauge-dependent.

To predict the existence of edge states between two lattices, we can still use the effective Hamiltonian or the topological invariant spin Chern number, but we need to ensure that the same gauge choice is made for both the connected lattices. This condition can be fulfilled by choosing the fundamental unit cells of the two lattices from an assembly such that they are separated by an integer number of lattice constants $\mathbf{a}_1, \mathbf{a}_2$.

\bibliography{REF_v3}